\documentclass[aps,prl,reprint,groupedaddress,showpacs]{revtex4-1}
\usepackage[utf8]{inputenc} \usepackage{verbatim}
\usepackage{xcolor,graphicx} \usepackage{rotate}
\usepackage{amsfonts,amssymb}

\begin{document}

\title{Asymptotic scaling behavior of self-avoiding walks on critical
  percolation clusters} \author{Niklas
  Fricke}\email[]{niklas.fricke@itp.uni-leipzig.de} \author{Wolfhard
  Janke}\email[]{wolfhard.janke@itp.uni-leipzig.de}

\affiliation{ Institut f\"ur Theoretische Physik and Centre for
  Theoretical Sciences (NTZ), \\ Universit\"at Leipzig, Postfach
  100920, D--04009 Leipzig, Germany }

\begin{abstract}
We study self-avoiding walks on three-dimensional critical percolation
clusters using a new exact enumeration method. It overcomes the
exponential increase in computation time by exploiting the clusters'
fractal nature. We enumerate walks of over $10^4$ steps, far more than
has ever been possible. The scaling exponent $\nu$ for the end-to-end
distance turns out to be smaller than previously thought and
{appears to be} the same on the backbones as on full
clusters.  We find strong evidence against the widely assumed scaling
law for the number of conformations and propose an alternative, which
perfectly fits our data.
\end{abstract}


\maketitle

The self-avoiding walk (SAW)~\cite{Madras1993} is a fundamental model
in statistical mechanics and crucial for our understanding of the
scaling behavior of polymers~\cite{DeGennes1976}. Asymptotically, it
is characterized by universal exponents, which are related to the
critical exponents of spin systems and assumed to describe long,
flexible polymers in good solvent condition. While much is known about
SAWs on regular lattices, their behavior in disordered environments,
such as porous rocks or biological cells, is less understood. The
paradigmatic model for such systems are SAWs on critical percolation
clusters~\cite{Chakrabarti2005,Ben-Avraham2000}. Here the walks can
only visit a random fraction of sites, whose concentration is equal to
the percolation threshold of the lattice. This critical concentration
may not be realistic, but it represents an important limiting case,
and the effect of the critical clusters' fractal structure is
particularly intriguing~\cite{Stauffer1992}.

One usually considers quenched disorder averages, here denoted by
square brackets: On each disorder realization (``cluster''), one takes
the average over all walk conformations of length $N$. Each such
conformational average contributes equally to the disorder average. It
is assumed that the average number of conformations, $\left[Z
  \right]$, and their mean squared end-to-end distance, $\left[ \left
  \langle R^2 \right \rangle \right ]$, follow asymptotic scaling laws
similar to those for normal SAWs:
\begin{eqnarray}
& \left[Z \right]& \sim \mu^N N^{\gamma-1}\label{ZN}, \\[1mm] & \left[
    \left \langle R^2 \right \rangle \right ] & \sim
  N^{2\nu}{=N^{2/d_f}}. \label{RN}
\end{eqnarray}
$\gamma$ and $\nu$ are universal scaling exponents,
{$d_f$ is the SAW's fractal (Hausdorff) dimension, }and
$\mu$ is a lattice dependent effective connectivity constant. While
the effect of the fractal disorder on $\gamma $ and $\mu $ is still
very controversial, there is convincing evidence that $\nu $ is
different than on regular lattices~\cite{Meir1989}. However, there is
uncertainty concerning the actual value despite a considerable amount
of work dedicated to the system. Analytical works have yielded
conflicting results~\cite{Ferber2004,Janssen2007,Janssen2012a}, while
accuracy and reliability of numerical investigations have been poor
due to modest system and sample sizes.  In most numerical studies (see
for instance
~\cite{Nakanishi1991,Vanderzande1992,Rintoul1994,Ordemann2000,Singh2009}),
exact enumeration was used to determine the conformational
averages. Owing to exponentially increasing computation time [see
  Eq.~(\ref{ZN})], the length of the walks was restricted to
$30-50$. Chain-growth Monte Carlo methods may allow for more than a
hundred
steps~\cite{Lee1988,Woo1991,Grassberger1993,Blavatska2008a,Blavatska2008},
but they add statistical uncertainty and the danger of biased
results~\cite{Fricke2013}.

We recently developed a new algorithm for exact enumeration of SAWs on
two-dimensional critical percolation clusters~\cite{Fricke2012a},
which we have now generalized to higher dimensions. By making use of
the clusters' fractal properties, it overcomes the exponential
increase in computation time that usually affects exact enumeration
methods. Walks of over $10^4$ steps are now accessible, permitting a
much more refined investigation of the system. Indeed, we think that
the true asymptotic behavior may now be revealed for the first time.

\begin{figure}[]
\centering \includegraphics[scale=0.21, trim= 0cm 0cm 0cm 0cm,
  clip=true]{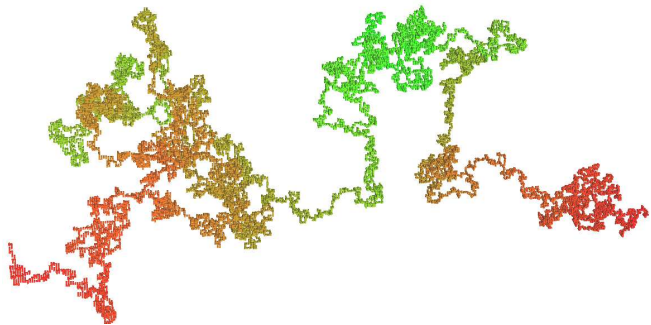}
\caption{Backbone of a critical percolation cluster on a cubic lattice
  of $300^3$ sites. The two ends are connected via periodic boundary
  conditions; coloring indicates shortest-path distance to the
  origin.}\label{f1}
\end{figure}

\begin{figure}[t!]
  \begin{center}
 \includegraphics[scale=0.52, trim= 6cm 10cm 9cm 2cm,
   clip=true]{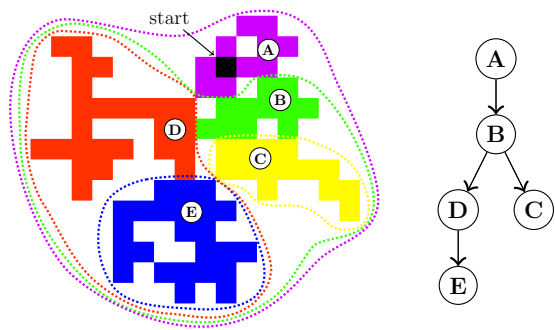}
  \end{center}
\caption{Schematic picture of a tree hierarchy of nested blobs. The
  starting position for the walks is marked black. Walk segments
  through the blobs are generated in the order $E\rightarrow
  A$. }\label{f2}
\end{figure}

Our method exploits the self-similar nature of the clusters to
factorize the problem hierarchically, drawing on the ideas of
renormalization group theory.  The key lies in the observation that
the connectivity of a critical percolation cluster is extremely low on
any length scale. This is best appreciated by looking at the backbone
of a cluster (Fig.~\ref{f1}), the part that remains when all
singly-connected ``dangling ends'' are removed. We define it as the
largest bi-connected component, i.e., the largest piece from which
nothing can be disconnected by removing a single site. The fragile
structure of the backbone, which is the most connected part, suggests
that the whole cluster can be decomposed into separate regions
by removing only a small number of sites~\cite{Coniglio1987}. This
applies on all length scales, so that we can organize the cluster into
a hierarchy of nested {``blobs''} as sketched in Fig.~\ref{f2}.  {Each
  blob should have very few ($\approx O(1)$) external connections and
  should not contain too much mass (colored areas in Fig.~\ref{f2})
  that is not encapsulated in smaller blobs. We create the blob
  hierarchy by repeatedly fusing regions with many interconnections,
  aiming for an optimal balance between the two requirements.}

The principle idea {now} is to factorize the enumeration procedure by
treating walk segments through different blobs separately. The
self-similarity of the system suggests a recursive approach: We start
by enumerating all possible walk conformations within smallest blobs
{(``children'')} that are contained within larger ones
{(``parents''). We use the standard backtracking
  routine~\cite{Nakanishi1991,Vanderzande1992,Rintoul1994,Ordemann2000,Singh2009}
  for this but distinguish different classes of paths depending on the
  the external connections they are linked to.  When we later generate
  the paths through the parents, the children are effectively treated
  as single sites, but we note by which links they are accessed and
  left. This information is needed to properly match the paths to the
  segments through the children once the counting within the parents
  is done. Now we can delete all information concerning the children
  and proceed on to the ``grand parents''.} This procedure is repeated
{up to} the largest blob, which is the whole cluster.

These main ideas are simple, but defining the blob hierarchy and
matching the path segments involves some technical challenges.  The
gain, however, is significant: On a present-day 3GHz processor, the
number of conformations for a $10^4$-step SAW (typically $10^{1200}$)
and their average end-to-end distance are determined in about 10
minutes on average using our current implementation. The exponential
complexity has vanished, and we empirically find a polynomial time
increase with an exponent around $2.4$. 

While the method works in any dimension, we here focus on the
physically most relevant case of $D=3$. To generate the clusters we
used a depth-first growth algorithm known as the Leath
method~\cite{Leath1976}. We only consider clusters that percolate
according to the ``wrapping'' criterion used in
Ref.~\cite{Newman2001}. The backbones were identified using Tarjan's
algorithm~\cite{Tarjan1974}. To avoid correlations, independent sets
of clusters were used for walks of different length, which we
increased by factors of $\sqrt{2}$ from $N=25$ to $N=12800$. Thanks
{to} the method's efficiency, we could afford samples of at least
$5\times 10^4$ clusters for each length.

\begin{figure}[t!]
\centering \includegraphics[scale=0.85, trim= 5.3cm 20.5cm 5cm 2cm,
  clip=true]{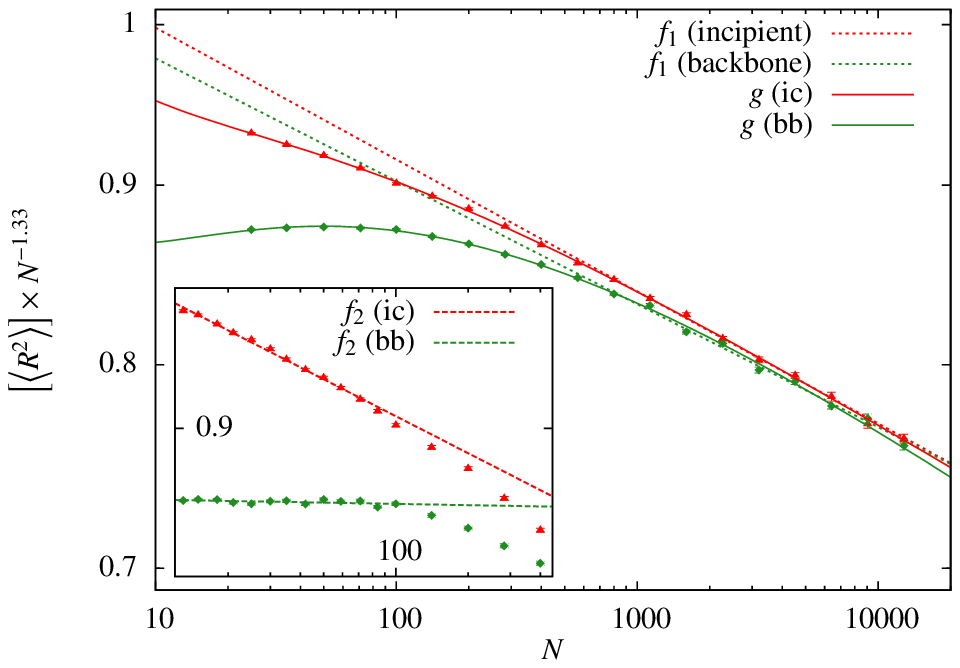}
\caption{Mean squared end-to-end distance vs number of steps for SAWs
  on incipient critical clusters (red) and backbones (green) on a
  log-log scale. The lines show the results from different
  least-square fits. $f_1$: Eq.~(\ref{RN}) with $N=800$--$12800$ (ic)
  and $N=1131$--$12800$ (bb); $f_2$ (inset): Eq.~(\ref{RN}),
  $N=13$--$100$; $g$: Eq.~(\ref{Rfit}), $N=25$--$12800$. The factor
  $N^{-1.33}$ ($\approx N^{-2\nu}$) serves to magnify the differences.
}\label{f3}
\end{figure}

The average squared end-to-end distance as a function of $N$ is shown
in Fig.~\ref{f3} on a double-logarithmic scale. To enhance
visibility, the values are divided by $N^{1.33}$, which is close to
$N^{2\nu}$ according to previous studies.  The curves appear straight
initially, but around $N\approx 150 $ they notably start to slump,
crossing over to a slightly different slope.  We hence have to use a
lower cutoff, $N_{\rm{min}}$, when estimating $\nu $ via a
least-squares fit of Eq.~(\ref{RN}).  On the whole (``incipient'')
clusters, the $\chi^2$-value of the fit becomes close to one
($\chi^2=1.3$) if we choose $N_{\rm{min}}=800$. This yields a value of
$\nu_{\rm{ic}}=0.6433(4)$. The fit is shown as dotted red line $f_1$
in Fig.~\ref{f3}. For the backbones, we get a decent fit
($\chi^2=3.4$) with $N_{\rm{min}}=1131$ (dotted green line), which
yields a similar result: $\nu_{bb}=0.643(1)$. Note, that the values of
$\chi^2 $ are meaningful since the data points are uncorrelated and
the errors are purely statistical.

It has often been
claimed~\cite{Rammal1984,Aharony1989,Ordemann2000,Blavatska2009a} that
the asymptotic statistics are determined by the backbone alone as a
dangling end can only support finite SAWs. As noted in
Ref.~\cite{Woo1991}, this argument is questionable since dangling ends
come in all sizes and have a larger fractal dimension than the
backbone. Still, our results provide strong evidence that
$\nu_{ic}=\nu_{bb}$ is indeed correct. As demonstrated in
Fig.~\ref{f3}, this becomes manifest only for sufficiently long
walks. The initial slope on the backbone is slightly larger, and the
asymptotic behavior is approached more slowly. This is somewhat
surprising: if the effects of the dangling ends vanish with $N$, it
should be the other way around, unless they happen to cancel out other
finite-size effects.

The fact that $\nu_{bb}$ appears larger initially explains why results
for $\nu$ from the most recent numerical studies, which only
considered the backbones, are significantly larger than ours
($0.662(6)$~\cite{Ordemann2000}, $0.667(3)$~\cite{Blavatska2008}). The
discrepancy is clearly due to the limited system sizes that had been
accessible ($30$ and $80$ steps, respectively). To verify this claim,
we used an upper cutoff of $N_{\rm{max}}=84$. Since calculations in this
regime are swift, we afforded a few more data points and increased the
sample sizes to $5\times 10^5$. As can be seen in the inset of
Fig.~\ref{f3}, a simple power-law nicely fits this data
($\chi_{bb}^2=2.0$, $\chi_{ic}^2=2.6$); one could hardly suspect a
different asymptotic behavior from this perspective. The resulting
backbone exponent $\nu_{bb}=0.6646(2)$ is consistent with previous
findings, though $\nu_{ic}=0.6547(2)$ is slightly (but significantly)
smaller.

Some of the finite-size effects may be explained by higher-order
corrections to Eq.~(\ref{RN}). Better fit results over a larger range
can indeed be obtained by including the next-to-leading confluent
correction term, $N^{2\nu-\Delta}$.  In practice, we fit
\begin{equation}
\left [ \left \langle R^2 \right \rangle \right ] = a (N+\delta
N)^{2\nu}\left (1+b/(N+\delta N)^{\Delta} \right ) \label{Rfit}
\end{equation}
as was done for the full-lattice SAW in Ref.~\cite{Clisby2010}. The
small shift, which we set to $\delta N=1/2$, provides for a smoother
convergence of the fit but has little effect on the actual
results. From the range $N=$ 25--12800 we thus obtain
$a_{ic}=1.13(2)$, $b_{ic}=-0.44(3)$; $a_{bb}=1.25(5)$,
$b_{bb}=-0.60(1)$ and
\begin{eqnarray}
\nu_{ic}=0.644(2), \quad \Delta_{ic}=0.51(5); \\ \nu_{bb}=0.640(3),
\quad \Delta_{bb}=0.34(4)
\end{eqnarray}
as our final estimates (for comparison,
$\nu_{\rm{full}}=0.587\;597(7)$ and $\Delta_{\rm{full}}=0.528(12)$
were found for the regular simple-cubic
lattice~\cite{Clisby2010}). The fits are shown as continuous curves
$g$ in Fig.~\ref{f3}. The $\chi^2$-values are $1.2$ (incipient
clusters) and $1.6$ (backbones). These estimates for $\nu$ are
consistent with those from the simple fits and again support
$\nu_{ic}=\nu_{bb}$. We also obtained similar (but less precise)
estimates by extrapolating the successive slopes, as was done
in~\cite{Lee1988,Rintoul1994,Ordemann2000}. {In terms
of the fractal dimensions, $d_f=1/\nu$, the results are
$d_{f,ic}=1.553(5)$, $d_{f,bb}=1.563(7)$.}
\begin{figure}[t!]
\centering \includegraphics[scale=0.87, trim= 2.5cm 20.5cm 5.8cm 2cm,
  clip=true]{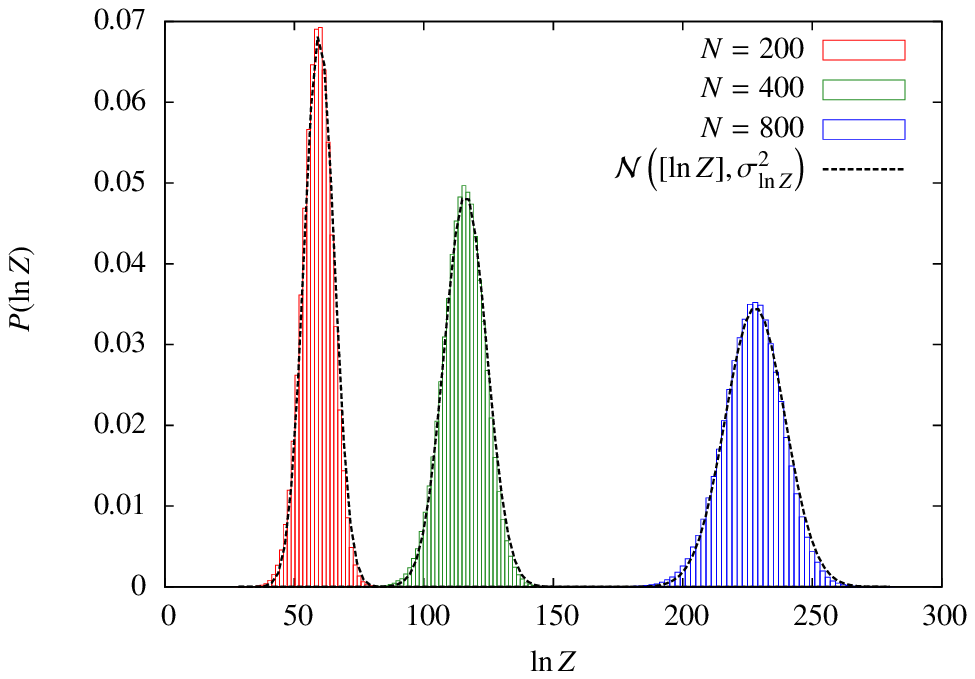}
\caption{Measured probability densities of the entropy $\ln{Z}$ for
  various lengths. The dashed lines are normal distributions with the
  same mean values and variances (no fit involved).}\label{f4}
\end{figure}

We now turn to the number of conformations, $Z$.  Here we only discuss
the results for the incipient clusters; those for the backbones are
qualitatively the same.  The distribution of $Z$ resembles a
log-normal as can be seen in Fig.~\ref{f4} where we have plotted the
measured frequencies of $\ln{Z}$ for various $N$ alongside normal
distributions with the same mean and variance. As noted
before~\cite{Grassberger1993}, such a ``multifractal'' distribution
can be explained by the fact that $Z$ is roughly a product of random
variables, namely the average coordination numbers at each
step. Indeed, we find for the variance of $\ln{Z}$:
\begin{equation}
\sigma^2_{\rm{\\ln{Z}}}\sim A N^{2 \chi }; \: A=0.1667(3), \:
\chi=0.500(1), \label{varlnZ}
\end{equation}
which supports this picture.  A similar result ($\chi=0.49(1)$) had
been reported previously~\cite{Rintoul1994}, but there appears to be
no theoretical explanation why $\chi=1/2$ should hold exactly.

These large deviations make it hard to obtain unbiased estimates for
$[Z]$: the value is easily underestimated if the sample size is too
small.  We managed to obtain reliable data for $N\leq200$ by pushing
the number of analyzed clusters to $10^7$. According to
Eq.~(\ref{ZN}), $\mu$ and $\gamma$ can be estimated by fitting
\begin{equation}
\ln{[Z]}/N = \ln{a}/N + \ln{\mu} + (\gamma-1)\ln{N}/N \label{lnZfit}
\end{equation}
as was done in~\cite{Ordemann2000, Blavatska2009a}. Trying this for
different ranges within $N\leq 200$, we found that the estimates for
$\mu$ ($\gamma$) systematically decrease (increase) with the upper
cutoff $N_{\rm{max}}$. This suggests that the asymptotic behavior is
not reached (which would not be surprising given our experience with
$\nu$). From our observations one might hence only infer the following
bounds:
\begin{equation}
\mu<1.440(4), \quad \gamma>1.9(1),
\end{equation}
obtained from a fit over $N=25$--$200$. However, such a large value
for $\gamma$ is very different from previous results (see, e.g.,
Table~4 in Ref.~\cite{Blavatska2009a}) and would be highly unusual.

\begin{figure}[]
\centering \includegraphics[scale=0.9, trim= 5.5cm 20.5cm 5.8cm 2cm,
  clip=true]{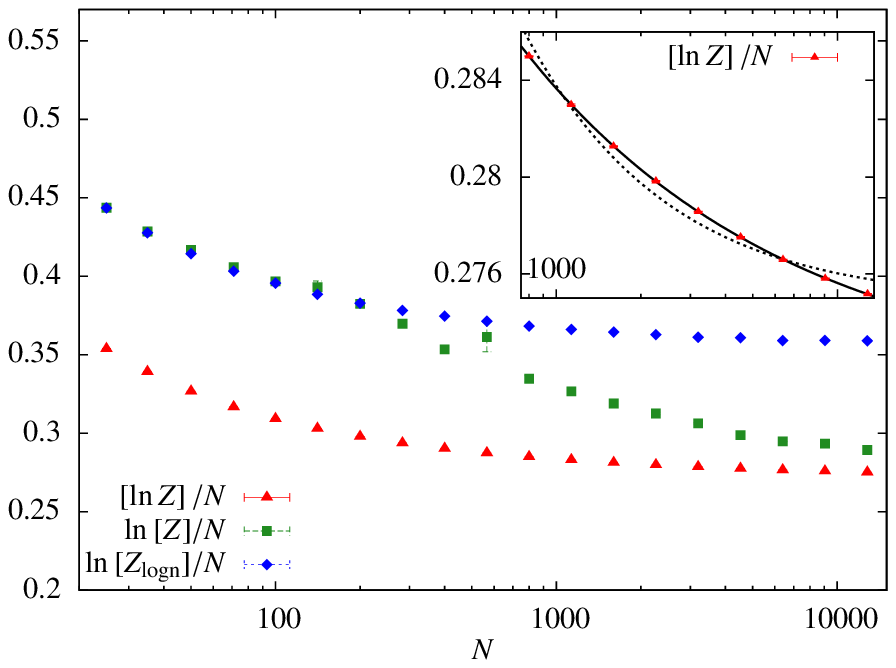}
\caption{Mean entropy (red triangles), logarithm of the average number
  of conformations (green squares), and log-normal approximation
  (blue diamonds) vs $N$ on a log-linear scale.  The estimates for
  $\ln{[Z]}$ are biased for $N>200$ due to large deviations. The inset
  shows fits of $[\ln{Z}]$ using Eq.~(\ref{lnZfit}) (dotted) and
  Eq.~(\ref{lnZfit2}) (continuous), respectively.}\label{f5}
\end{figure}

For longer chains, we can only approximate $[Z]$ by assuming the
distribution of $Z$ to be log-normal:
\begin{equation}
\left [ Z \right ] \approx
e^{[\ln{Z}]+\sigma^2_{\ln{Z}}/2}=\mathrel{\mathop:} \left [
  Z_{\rm{logn}} \right].\label{Z_lognorm}
\end{equation}
$\left[ Z_{\rm{logn}} \right]$ can be estimated more easily since the
``entropies'', $\ln{Z}$, are better behaved.  In Fig.~\ref{f5} we
have plotted $\left [ \ln{Z} \right ]/N$, $\ln{\left [Z \right ]}/N $,
and $\ln{\left[ Z_{\rm{logn}} \right] }/N$ vs $N$. As can be seen,
$[Z]\approx \left[ Z_{\rm{logn}} \right]$ is fulfilled well for small
$N$. For larger $N$, $\ln{\left [ Z \right ] }$ appears to approach
$[\ln{Z}]$, which is a consequence of the aforementioned bias.

The exponential of the mean entropy,
$[Z_0]\mathrel{\mathop:}=e^{[\ln{Z}]}$, is supposed to follow a
scaling law similar to \mbox{Eq.~(\ref{ZN})~\cite{Ordemann2000}}. Here
we have reliable data up to $N=12800$, which should in principle allow
for accurate estimates of the ``zeroth moments'', $\mu_0$ and
$\gamma_0$. However, a fit analogous to Eq.~(\ref{lnZfit}) for
$[\ln{Z}]/N$ yields poor results, which remain strongly dependent on
the fit range. (We made a similar observation for the two-dimensional
case~\cite{Fricke2012a}.) The data are much better described by a
function of the form
\begin{equation}
\left[\ln{Z}\right]/N = \ln{a}/N +
{\left (\ln{\mu_0} \right ) \left (1+bN^{-\zeta} \right )} \label{lnZfit2}
\end{equation}
as can be seen in the inset of Fig.~\ref{f5}. Using
Eq.~(\ref{lnZfit2}), the fit results stabilize around
$N_{\rm{min}}\approx 800$, which is consistent with the findings for
$\nu$. From the range $N=800$--$12800$ we obtain: $a=0.7(4)$,
$\ln{\mu_0}=0.2715(3)$, $\zeta=0.48(3)$, and $b=1.3(3)$ with
$\chi^2=0.52$. This would imply:
\begin{equation} 
[Z_0]\sim \mu_0^{N(1+b/N^{\zeta})} \quad \rm{with} \quad
\mu_0=1.3119(3), \label{Z0}
\end{equation}
rather than a scaling law of the form of Eq.~(\ref{ZN}).  There might
still be a factor $N^{\gamma_0-1}$, but we found no numerical evidence
for it.  Unfortunately, we cannot do a similar fit for $\ln{[Z]}/N$
for lack of reliable data points. However, if Eq.~(\ref{Z0}) is
correct for $[Z_0]$ and assuming Eq.~(\ref{varlnZ}), we can infer a
similar law for $\left[ Z_{\rm{logn}} \right]$ with $\mu_{\rm{logn}}=
e^{\ln{\mu_0}+A/2} = 1.4260(6)$.  $[Z]\approx \left[ Z_{\rm{logn}}
  \right] $ would then suggest a scaling law like Eq.~(\ref{lnZfit2})
for $[Z]$ as well. That approximation is not well-founded, so
Eq.~(\ref{ZN}) might still be correct, but we think that the empirical
evidence against it is significant.  {In fact, there is
  also little theoretical foundation for Eq.~(\ref{ZN}) other than the
  analogy to the regular-lattice case where numerical and analytical
  support for such a scaling law is strong.}  The unusual correction
term {in Eq.~(\ref{Z0})} may arise from the
non-self-averaging properties of the critical clusters. A similar law,
but with $b<0$, has been found for conformations of random walks on
percolation clusters~\cite{Giacometti1994a}.

In any case, our results clearly disprove that $\mu$ results as the
undiluted value times the critical concentration, $\mu\approx p_c
\mu_{\rm{full}}=1.45958(2)$~\cite{Xu2014,Schram2011}, claimed
in~\cite{Woo1991,Ordemann2000,Blavatska2009a}. By restricting the
range of $N$ we again get a similar value, which is probably due to
finite-size effects: Initially, the lattice defects and the
self-avoidance act independently; only with increasing $N$ does their
interplay and the topology of the clusters become relevant. The
asymptotic behavior might, for instance, be affected by the
distribution of loop sizes on the cluster (backbone), or by the
spatial distribution of regions that contribute disproportionately to
the entropy, which cannot be gauged by short walks.

In summary, we have presented a method to exactly enumerate SAWs of
over $10^4$ steps on three-dimensional critical percolation
clusters. This enabled a firm analysis of the asymptotic scaling
behavior of the end-to-end distance with unprecedented accuracy.  We
revised the established estimate for the leading scaling exponent,
verified the hypothesis $\nu_{ic}=\nu_{bb}$, and gave a first estimate
for the confluent correction exponent $\Delta$.

Direct investigation of the average number of conformations, $[Z]$,
was hampered by large deviations, rendering our results less
conclusive here. The nature of the distribution of $\ln{Z}$, which
resembles a Gaussian whose variance we found to increase linearly with
$N$, suggests that information can be gleaned from the mean entropy,
$[\ln{Z}]$.  Surprisingly, $[\ln{Z}]$ does not behave as expected,
which puts the commonly assumed scaling law for $Z$ [Eq.~(\ref{ZN})]
into question as well.

Our findings show that the true asymptotic scaling behavior cannot be
observed from system sizes accessible with other numerical tools. This
observation may serve as a general lesson of caution regarding
numerical studies of systems with strong (fractal) disorder, and calls
for further applications of our new method. These may include other
types of walks or media (or both). The SAW can be furnished with
short-range interactions to model
$\Theta$-polymers~\cite{Roy1990,Barat1995}, possibly under stretching
force~\cite{Blavatska2009, Singh2009}. One can also add bending
stiffness to study semi-flexible
polymers~\cite{Giacometti1992,Lekic2011}. The underlying idea of a
scale-free partitioning is not even restricted to walk models but
could be transferred to spin systems or transport processes. Of
course, the necessary condition is that the medium has a weakly
connected, self-similar geometry. One can obviously study percolation
clusters of different dimensionality, even beyond the upper critical
dimension of $D=6$, to gain deeper understanding of the role of the
medium's fractal dimensions. For $p>p_c$, the efficiency of our method
eventually deteriorates, but it can still beat other methods near the
critical concentration~\cite{Fricke2013}. Further applications could
include Ising and Potts clusters, DLA clusters, and possibly certain
types of quantum gravity graphs~\cite{Janke2000} or real-world fractal
networks~\cite{Song2004}.

\begin{acknowledgments}
This work was funded by the Deutsche Forschungsgemeinschaft (DFG) via
FOR 877, Grant No.\ JA 483/29-1, and SFB/TRR 102 (project B04). We are
grateful for further support from Graduate School GSC 185
``BuildMoNa'', Deutsch-Franz\"osische Hochschule (DFH) under Grant
No.\ CDFA-02-07, and an AvH Institute Partnership Grant with Lviv,
Ukraine.
\end{acknowledgments}

\bibliography{references.bib}

\end{document}